\documentclass[reprint, prd, superscriptaddress, longbibliography, tightenlines, nofootinbib, eqsecnum, amsfonts, amsmath, amssymb, floatfix, notitlepage, twocolumn]{revtex4-1}
\pdfoutput=1

\usepackage[utf8]{inputenc}
\usepackage{graphicx}
\usepackage[usenames,dvipsnames,svgnames,table]{xcolor}
\usepackage{xspace}
\usepackage[nolist,nohyperlinks]{acronym}
\usepackage{hyperref}
\usepackage{silence}
\graphicspath{{Figures/}}

\newcommand{\pMAP}[0]{\ensuremath{\phi^{\rm MAP}}\xspace}

\newcommand{\hpMAP}[0]{\ensuremath{\hat \phi^{\rm MAP}}\xspace}
\newcommand{\hpit}[0]{\ensuremath{\hat \phi^{\rm it}}\xspace}

\newcommand{\hpQE}[0]{\ensuremath{\hat \phi^{\rm QE}}\xspace}

\newcommand{\pMFone}[0]{\ensuremath{\phi^{\rm MF, 1}}\xspace}
\newcommand{\pMFtwo}[0]{\ensuremath{\phi^{\rm MF, 2}}\xspace}
\newcommand{\Xdat}[0]{\ensuremath{ {X^{\rm dat}}}\xspace}% CMB data
\newcommand{\Xdatg}[0]{\ensuremath{ {X^{\rm dat, \dagger}}}\xspace}% CMB data
\newcommand{\cppfid}[0]{\ensuremath{C^{\phi\phi, \rm fid}_L}\xspace}
\newcommand{\cppin}[0]{\ensuremath{C_L^{\phi \phi, \rm in}}\xspace}
\newcommand{\cppit}[0]{\ensuremath{C_L^{\phi \phi, \rm it}}\xspace}
\newcommand{\cpptrue}[0]{\ensuremath{C_L^{\phi \phi, \rm true}}\xspace}
\newcommand{\fsky}[0]{\ensuremath{f_{\rm sky}}\xspace}
\newcommand{\cpp}[0]{\ensuremath{C^{\phi\phi}_L}\xspace}
\newcommand{\Cov}[0]{\ensuremath{\textrm{Cov}}\xspace} % Covariance matrix
\newcommand{\LM}[0]{{LM}}

\newcommand{\av}[1]{\left \langle #1 \right \rangle}

\newcommand{\RDNLZERO}{\ensuremath{\mathrm{RD{\text -}}N_L^{(0)}}\xspace}
\newcommand{\MCNLONE}{\ensuremath{\mathrm{MC{\text -}}N_L^{(1)}}\xspace}
\newcommand{\Nlone}{\ensuremath{N_L^{(1)}}\xspace}
\newcommand{\Nlzero}{\ensuremath{N_L^{(0)}}\xspace}

\newcommand{\Eunl}[0]{\ensuremath{E^{\rm unl}}} % Unlensed E-mode
\newcommand{\CEEunl}[0]{\ensuremath{C^{EE,\rm unl}}\xspace} % Unlensed E-mode spectrum
\newcommand{\CEEunli}[0]{\ensuremath{C^{EE,\rm unl, -1}}} % Unlensed E-mode spectrum inverse matrix
\newcommand{\Y}[0]{ \ensuremath{\: _{2}\mathcal Y}}

\newcommand{\Beam}[0]{\ensuremath{\mathcal B}\xspace}
\newcommand{\Da}[0]{\ensuremath{\mathcal D_{\va}}\xspace} % deflection operator at alpha
\newcommand{\Punl}{\ensuremath{\: _{2}P^{\rm unl}}\xspace} % unlensed polarization.
\renewcommand{\a}[0]{\ensuremath{\alpha}} % letter. Aim is to have to change this one only
\newcommand{\va}[0]{\ensuremath{\boldsymbol{\a}}} % vector field
\newcommand{\Na}[0]{\ensuremath{N_{\va}}}

\newcommand{\Cova}{\ensuremath{\textrm{Cov}_{\va}}} % data.
\newcommand{\barStt}{{\bar P}_{\va}} % Inverse-variance weighted leg
\newcommand{\Stwf}{ P^{\rm WF}_{\va}} % Wiener-filtered leg

\newcommand{\barSttqe}{{\bar P}} % Inverse-variance weighted leg
\newcommand{\Stwfqe}{ P^{\rm WF}} % Wiener-filtered leg

\newcommand{\EWFlm}{\ensuremath{E^{\rm WF}_{\va, \ell m}}} % Wiener-filtered E
\newcommand{\EWF}{\ensuremath{E^{\rm WF}_{\va}}} % Wiener-filtered E
\newcommand{\curlpot}[0]{\ensuremath{\Omega}}

\newcommand{\vn}[0]{\ensuremath{\boldsymbol{n}}} 
\newcommand{\chis}[0]{\ensuremath{\chi^2}\xspace} 

\newcommand{\Planck}{\textit{Planck}\xspace}

\begin{document}

\title{Robust and efficient CMB lensing power spectrum from polarization surveys}

\author{Louis Legrand }\email{louis.legrand@unige.ch}
\affiliation{Universit\'e de Gen\`eve, D\'epartement de Physique Th\'eorique et CAP, 24 Quai Ansermet, CH-1211 Gen\`eve 4, Switzerland}
\affiliation{ICTP South American Institute for Fundamental Research, Instituto de F\'{\i}sica Te\'orica, Universidade Estadual Paulista, S\~ao Paulo, Brazil}

\author{Julien Carron}\email{julien.carron@unige.ch}
\affiliation{Universit\'e de Gen\`eve, D\'epartement de Physique Th\'eorique et CAP, 24 Quai Ansermet, CH-1211 Gen\`eve 4, Switzerland}

\date{\today}

\begin{abstract}
    Deep surveys of the CMB polarization have more information on the lensing signal than the quadratic estimators (QE) can capture. We showed in a recent work that a CMB lensing power spectrum built from a single optimized CMB lensing mass map, working in close analogy to state-of-the-art QE techniques, can result in an essentially optimal spectrum estimator at reasonable numerical cost. We extend this analysis here to account for real-life non-idealities including masking and realistic instrumental noise maps. As in the QE case, it is necessary to include small corrections to account for the estimator response to these anisotropies, which we demonstrate can be estimated easily from simulations. The realization-dependent debiasing of the spectrum remains robust, allowing unbiased recovery of the band powers even in cases where the statistical model used for the lensing map reconstruction is grossly wrong. This allows robust and optimal CMB lensing constraints from CMB data, on all scales relevant for the inference of the neutrino mass, or other parameters of our cosmological model.
\end{abstract}

\maketitle

\section{Introduction}

The gravitational lensing of the Cosmic Microwave Background is a comparatively clean and robust probe  of the large scale structures of our Universe \cite{Lewis:2006fu}. Accurate and precise measurement of this signal is a key target for most upcoming CMB surveys such as SPT-3G, Simons Observatory and CMB-S4 \cite{SPT-3G:2021vps, SimonsObservatory:2018koc,CMB-S4:2016ple}. On the one hand, CMB lensing mass maps must be used to reduce the sample variance of the observed polarization B modes in order to constrain inflation \cite{Seljak:2003pn, Carron:2017vfg, Namikawa:2021gyh}, can serve to calibrate the mass of galaxy clusters at high redshifts \cite{Lewis:2005fq,Hu:2007bt, Raghunathan:2017cle, DES:2018myw}, and will cross-correlate to other probes of large-scale structure such as galaxy surveys. On the other hand, the CMB lensing power spectrum promises tight constraints on the sum of neutrino masses~\cite{Kaplinghat:2003bh,dePutter:2009kn,Allison:2015qca}.

The quadratic estimator (QE) \cite{Hu:2001kj,Okamoto:2003zw, Maniyar:2021msb} has been the standard tool to reconstruct the CMB lensing potential until now \cite{Planck:2018lbu,Wu:2019hek,Darwish:2020fwf}.
CMB lensing creates anisotropies in the two point function of the observed CMB fields. The QE reconstructs the lensing field by using the mode coupling created by lensing in pairs of CMB maps. However, the QE will not be optimal for next generation CMB surveys, as its variance is limited by the lensed CMB spectra. Surveys such as CMB-S4 will have instrumental sensitivity well below the lensing induced $B$ mode power of $\sim 5 \mu \rm{K} \rm \:arcmin$ over wide areas of the sky. At the small scales relevant for lensing reconstruction, we can neglect the primordial $B$ modes, sourced by the gravitational waves from inflation. Likelihood-based estimators \cite{Hirata:2003ka} can then reconstruct optimally the lensing field from the observed lensed $E$ and $B$ modes map. This allows to greatly decrease the lensing reconstruction variance, by being limited by the delensed data spectra instead of the lensed spectra.

Several methods have been proposed to reconstruct the lensing power spectrum~\cite{Hirata:2003ka, Carron:2017mqf, Millea:2017fyd,Millea:2020cpw,Millea:2021had, Chan:2023vye} beyond the QE.
In a previous work~\cite{Legrand:2021qdu}, we showed that proper characterization of the response and biases to the maximum-a-posteriori lensing map~\cite{Hirata:2003ka, Carron:2017mqf} spectrum gives an unbiased lensing power spectrum in a manner that closely follows the standard procedure for QEs. We see three advantages to that approach. First, the approach requires the reconstruction of a single lensing map, and is thus very economical compared to other possibilities such as sampling~\cite{Millea:2020cpw}, or iterative spectrum reconstruction~\cite{Hirata:2003ka, Millea:2021had}. This allows spectrum reconstruction on large sky fractions at moderate numerical cost. Second, this approach achieves the expected best possible signal to noise on the lensing spectrum, as calculated first by Ref.~\cite{Smith:2010gu}, hence is also essentially optimal. Third, in analogy to the QE, the spectrum can be cleaned of the leading noise bias term in a realization-dependent manner, hence is robust to mischaracterization of the data statistical properties, which is quite important in practice.

The analysis of~\cite{Legrand:2021qdu} was performed on idealized, full-sky simulations, with isotropic Gaussian noise.
In order to use the optimal lensing estimator for an actual data-analysis, we need to evaluate and mitigate the impact of real-life non-idealities, such as the mask of the survey, or the presence of anisotropic and non-Gaussian noise.
This is the scope of the present paper: we estimate the different corrections on the iterative lensing responses due to the mask, and show that they essentially depend only on fiducial ingredients. Hence they can be calibrated, similarly to the procedure applied for the QE. We also show that we recover unbiased lensing power spectrum in the presence of non-Gaussian anisotropic noise.

We start in Sec.~\ref{sec:reconstruction} by reviewing iterative, maximum-a-posteriori lensing reconstruction. We then discuss in Sec.~\ref{sec:mf} the mean-field contamination to the iterative lensing map created by the mask and our way of treating it. We then estimate corrections on the lensing power spectrum normalization in Sec~\ref{sec:norm}. We adapt in Sec.~\ref{sec:clpp}  the realization-dependent noise debiaser to these realistic cases. Finally, in Sec.~\ref{sec:nonfidsim} we perform a series of test spectrum reconstructions with varying inputs skies and non-idealities and show that we recover the expected signal.

We follow the standard convention and denote lensing multipoles with $(L, M)$ and CMB multipoles with $(l, m)$.

\section{Lensing reconstruction}\label{sec:reconstruction}

We perform lensing reconstruction based on the polarization maps only, neglecting the signal that can be extracted from the temperature map. For CMB-S4, the polarization only estimator will dominate the lensing reconstruction signal \cite{CMB-S4:2016ple}. Moreover, the polarized CMB maps have the advantage of being less contaminated by foreground signals compared to the temperature maps \cite{Fantaye:2012ha,Beck:2020dhe,Sailer:2022jwt}.

The curved sky polarization is expressed with the spin-2 polarization field, defined from the Stokes parameters $Q$ and $U$, which can be related to the spherical harmonics decomposition $E$ and $B$ modes
\begin{equation}
    _{\pm 2} P (\vn) \equiv Q(\vn) \pm i U(\vn) = - \sum_{lm} (E_{lm} \pm i B_{lm} )\,  _{\pm 2} Y_{lm}(\vn) \, ,
\end{equation}

We will use a compact notation for the spherical harmonic transformation of the unlensed fields $\Punl = \Y \Eunl$.
We neglect the unlensed $B$ modes. In the small scale regime relevant for the lensing reconstruction, the potential primordial $B$ modes created by primordial gravitational waves would be dominated by the lensing induced $B$ modes.

Let \Da be the deflection operator which maps the unlensed polarization field \Punl into its lensed components. This deflection operation can be expressed as a remapping of the polarization fields, under the deflection vector \va, with a slight phase shift (see \cite{Carron:2017mqf, Belkner:2023} for the full expression).
The deflection spin-1 vector field can be expressed as the sum of a gradient and curl components. 
\begin{equation}
        _{1}\alpha(\hat n) = - \eth \phi(\hat n) - i \eth \curlpot(\hat n) \; ,
\end{equation}
where $\eth $ is the spin raising operator \cite{Lewis:2001hp}. 
We will neglect the small curl component here \cite{Hirata:2003ka, Robertson:2023xkg}.

The observed Stokes fields can be expressed as 
\begin{equation}
    \label{eq:xdat}
    \Xdat \equiv \begin{pmatrix} Q^{\rm dat } \\ U^{\rm dat } \end{pmatrix} =\Beam \Da \Punl + \textrm{noise}
\end{equation}
where \Beam is the operator containing the instrument beam and transfer functions, as well as the projection of the lensed polarization spin-2 field into its $Q$ and $U$ components.

The covariance of the observed CMB fields, under a fixed lensing field \va, is given by 
\begin{equation}
    \begin{split}\label{eq:cov}
        \Cova &\equiv \av{\Xdat \Xdatg}_{\va}  \\ &= \Beam \Da \Y \:\CEEunl \Y^\dagger \Da^\dagger \Beam^\dagger + N \, .
    \end{split}
\end{equation}
where \CEEunl is the fiducial unlensed $E$ modes spectra. We assume that the polarization noise is independent from the signal and is described by the pixel covariance matrix $N$.

The likelihood of the lensing field given the observations is
\begin{equation}
    \label{eq:loglike}
    \ln \mathcal{L}( \Xdat | \va) = - \frac{1}{2}\Xdat^\dag \Cov_{\va}^{-1}\Xdat - \frac{1}{2} \ln \det \Cov_{\va} \;.
 \end{equation}
Putting a Gaussian prior on the lensing field, we can write the posterior
\begin{equation}
    \label{eq:logpost}
        \ln \mathcal{P}(\va | \Xdat) = \ln \mathcal{L}( \Xdat | \va) - \frac{1}{2}\sum_{\LM}  \frac{\phi_{\LM} \phi_{LM}^\dagger}{\cppfid}  \; .   
\end{equation}

The maximum a posteriori (MAP) lensing field \pMAP is found by maximizing Eq.~\ref{eq:logpost}. In practice, following \cite{Carron:2017mqf}, this MAP is found with a Newton-Raphson iterative procedure. 
At each iteration, the next lensing field estimate is obtained by computing the gradient of the posterior with respect to the lensing field. 
This gradient is a spin-1 field, and can be decomposed into three terms
\begin{equation}
    \label{eq:gradtot}
    \begin{split}
        \:_{\pm 1}{g_{\va}^{\rm tot}}(\vn) &= e^a_{\pm} \frac{\delta }{\delta \alpha^a(\vn)}\left[ \ln \mathcal P (\va | \Xdat )\right] \\
         &= \:_{\pm 1}{g_{\va}^{\rm QD}}(\vn) - \:_{\pm 1}{g_{\va}^{\rm MF}}(\vn) + \:_{\pm 1}{g_{\va}^{\rm PR}}(\vn)  \, , 
    \end{split}
\end{equation}
with $e^a_\pm$ the spin basis associated to the spherical coordinates basis vectors \cite{Lewis:2001hp}.

The quadratic piece of the likelihood gradient can be written in real space as the product of two maps (each \textit{leg} of the estimator), one is inverse variance filtered (IVF)  $\barStt$  and the other Wiener-filtered (WF) $\Stwf$ 
\begin{equation}
    \label{eq:QDgradient}
    \:_{1}{g_{\va}^{\rm QD}}(\vn) = -\sum_{s=\pm 2} \:_{-s}{\barStt}(\vn)\left[\Da\:\eth_s \Stwf\right](\vn) \, ,
\end{equation}
The expressions of the IVF and WF fields are given by  (see \cite{Carron:2017mqf, Belkner:2023} for a complete derivation)
\begin{equation}
     \begin{split}
        _2 \barStt(\vn) &\equiv \left[\mathcal B^\dagger \Cova^{-1} \Xdat \right] (\vn)\\
         &= \left[ \mathcal B^\dagger N^{-1}\left(  \Xdat - \mathcal B \Da \Y \EWFlm \right)\right](\vn) \\ 
    \end{split}
\end{equation}
The WF $E$ mode is
\begin{equation}
        \label{eq:ewf}
        \EWF \equiv \left[\CEEunli + \Na^{-1}\right]^{-1}  \Y^\dagger \Da^\dagger \mathcal B^\dagger N^{-1}\Xdat
\end{equation}
where $\Na$ is the delensed $E$-noise covariance matrix, 
\begin{equation}
    \label{eq:delensedNoise}
    \Na^{-1} \equiv \Y^\dagger\Da^\dagger \mathcal B^\dagger  N^{-1} \mathcal B \Da \Y.
\end{equation}
Because we assume there are no primordial $B$ modes, there are no Wiener-filtered $B$ modes either.

The inversion inside the brackets of Eq.~\ref{eq:ewf} is performed with a conjugate gradient descent. The noise matrix $N$ has a diagonal component in pixel space, corresponding to the fiducial noise statistical model. The pixels that fall in the mask are given an infinite variance.

The gradient of the determinant term of the likelihood is analogous to a mean-field. It can be expressed as the average of the quadratic gradients over realization of the data statistical model, for fixed deflection field $\va$:$\:_{1}{g_{\va}^{\rm MF}}(\vn) = \left< \:_{1}{g_{\va}^{\rm QD}}(\vn) \right>_{\va}$. 
Part of this mean-field is sourced by the delensed noise matrix of Eq.~\ref{eq:delensedNoise}. Indeed, even if the observed noise matrix were isotropic, the delensing of the noise will create anisotropies by compressing or dilating the noise according to the local convergence. This leaves a mean-field signal when averaging fixing the deflection field.
As such this component is still present even for a full-sky reconstruction, in contrast to the QE case. However, this delensing induced mean-field is small, and can be neglected in the full-sky reconstruction. 

It would be too computationally expensive to run a set of Monte-Carlo (MC) simulations at each iteration in order to estimate the full mean-field. We discuss our treatment in more detail in Section~\ref{sec:mf}.

The quadratic gradient of Eq.~\ref{eq:QDgradient} is the analogous of the standard QE written in real space, which we also calculate for comparison. The main difference is that the QE does not express the estimate of the lensing operation in terms of the covariance matrix of the CMB fields
\begin{equation}
    \begin{split}
    \Cov &\equiv \av{\Xdat \Xdatg}  \\ &= \Beam \Y C^{\rm fid}  \Y^\dagger \Beam^\dagger + N \, .
    \end{split}
\end{equation}
where $C^{\rm fid} $ is a matrix containing the fiducial lensed CMB spectra $EE$ and $BB$.
We can write the standard polarization only QE as 
\begin{equation}
    \label{eq:QEgradient}
    \:_{1}{g^{\rm QE}}(\vn) = -\sum_{s=\pm 2} \:_{-s}{\barSttqe}(\vn)\left[\eth_s \Stwfqe \right](\vn) \, ,
\end{equation}
where the two legs are
\begin{equation}
    \begin{split}
        \:_2 \barSttqe(\vn) &\equiv \left[\mathcal B^\dagger \Cov^{-1} \Xdat  \right](\vn)\\
        \:_2\Stwfqe(\vn) &\equiv \left[ C^{\rm fid}  \Y^\dagger \mathcal B^\dagger \Cov^{-1}\Xdat \right] (\vn)\\
    \end{split}
\end{equation}
The inversion of the covariance matrix is performed with a conjugate gradient descent as well, similarly to the MAP expression.

In both cases we reconstruct the lensing potential fields for $2\leq L \leq 4000$
\footnote{Our reasoning to omit $L=1$ is as follows: in the early stages of this work, we found that our curved-sky lensing reconstruction algorithm had some convergence issues at the lowest lensing multipoles. This problem worsened in the presence of aberration, the very large, essentially pure dipole, lensing signature resulting from our motion with respect to the CMB frame~\cite{Planck:2013kqc}. This issue is due to the necessity of precise lensing and delensing remapping implementations to reconstruct the lowest $L$'s, and this was difficult to achieve at a reasonable numerical cost. 
For this reason we used in early stages of this work simulations without any lensing signal at $L=1$. This issue was completely resolved in the meantime, thanks to major improvements in accuracy and efficiency of our lensing codes \cite{Reinecke:2013,Reinecke:2023gtp}. However, for the sake of self-consistency we continued to employ these lensing dipole-free simulations while finalizing this paper.} 
using the CMB multipoles in the range $2\leq \ell \leq 3000$. Reconstruction of the delensed $E$-mode in Eq.~\eqref{eq:ewf} occurs for $2\leq \ell \leq 4000$.
We consider a full-sky polarization survey, with a fiducial isotropic CMB-S4 like noise level of $ \sqrt{2} \, \rm \mu K\:arcmin$, and a beam full width at half maximum of $1 \, \rm arcmin$. 
\section{Mean-field}
\label{sec:mf}
\begin{figure}
    \centering
    \includegraphics[width=\columnwidth]{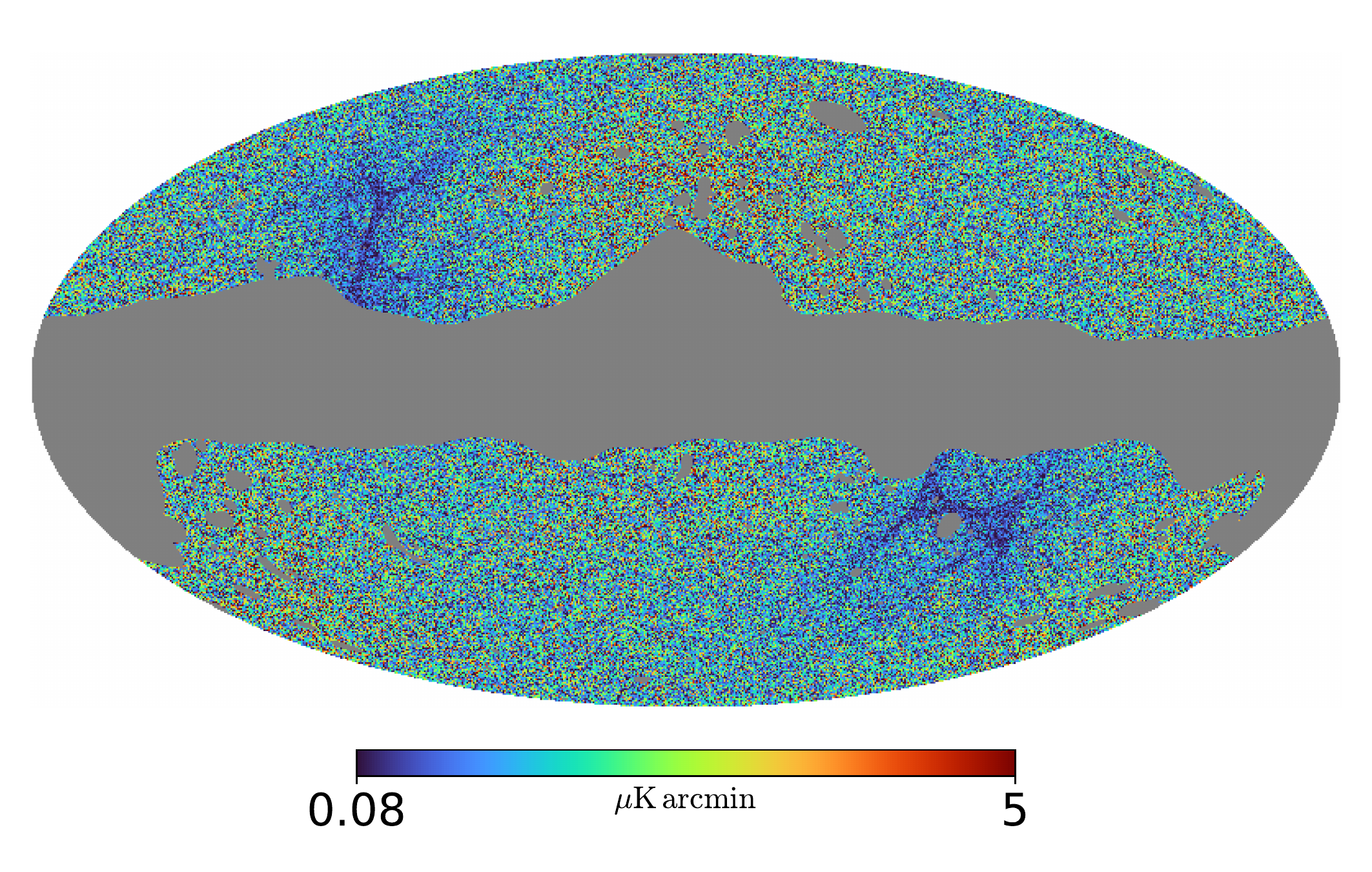}
    \caption{Map of a simulated anisotropic noise in polarization, given as the residual signal contained in one of the \Planck FFP10 simulations after subtraction of the estimated CMB and foregrounds signals. We plot here $\sqrt{Q^2 + U^2}$ in $\mu\rm{K}\:arcmin$. We have multiplied the polarization maps by a constant to obtain a full-sky variance of $\sqrt{2} \rm \, \mu \rm{K}\: arcmin$. A small amount of pixels are below or above the threshold of the colorbar, which shows the dynamical range of crudely one order of magnitude. The grey area corresponds to the \Planck lensing mask, which we use as test case for our masked reconstructions.}
    \label{fig:anisonoise}
\end{figure}
We use the \Planck lensing mask, shown in Fig.~\ref{fig:anisonoise}, to perform reconstructions on masked data. One reason for this choice is that the QE reconstructions on this mask have been intensively studied through the series of \Planck lensing releases~\cite{Planck:2013mth, Planck:2015mym,Planck:2018lbu, Carron:2022eyg}, inclusive of the origin of the spectrum Monte-Carlo correction~\cite{Carron:2022edh}, which cannot be modelled analytically. The full mask consists mainly of a mixture of a galactic mask with selected point source and cluster masks, with an unmasked sky fraction of $67 \%$, a number crudely in line with the CMB-S4 wide survey current specification.

In \cite{Legrand:2021qdu}, we used a perturbative approximation introduced in \cite{Carron:2017mqf} to estimate the variation of the lensing induced mean-field between two iterations. This perturbative approach was found to be accurate in the absence of non-idealities, i.e. when the only source of the mean-field is the delensed noise anisotropy. However, we found that in the presence of masking, this perturbative approximation (valid only away from the mask boundaries) was having an adverse effect near the mask causing difficulties in the MAP search.

In practice we found that it is satisfactory to neglect the mean-field term that is induced by lensing when performing the iterations. We take $ g^{\rm MF} _{\va} = g^{\rm QE, MF} $, the QE mean-field at all iterations. 
Once the \hpMAP is converged, we estimate the residual mean-field.
To estimate this residual mean-field we proceed as for a standard QE. We use a set of simulations, with independent CMB and lensing field realization, but with the same mask. We reconstruct \hpMAP on each of them. The residual mean-field is then estimated as the average of these \hpMAP estimates. 
This neglects the delensed-noise mean-field. As shown below, we found that this approximation is satisfactory.

\begin{figure}
    \centering
    \includegraphics[width=\columnwidth]{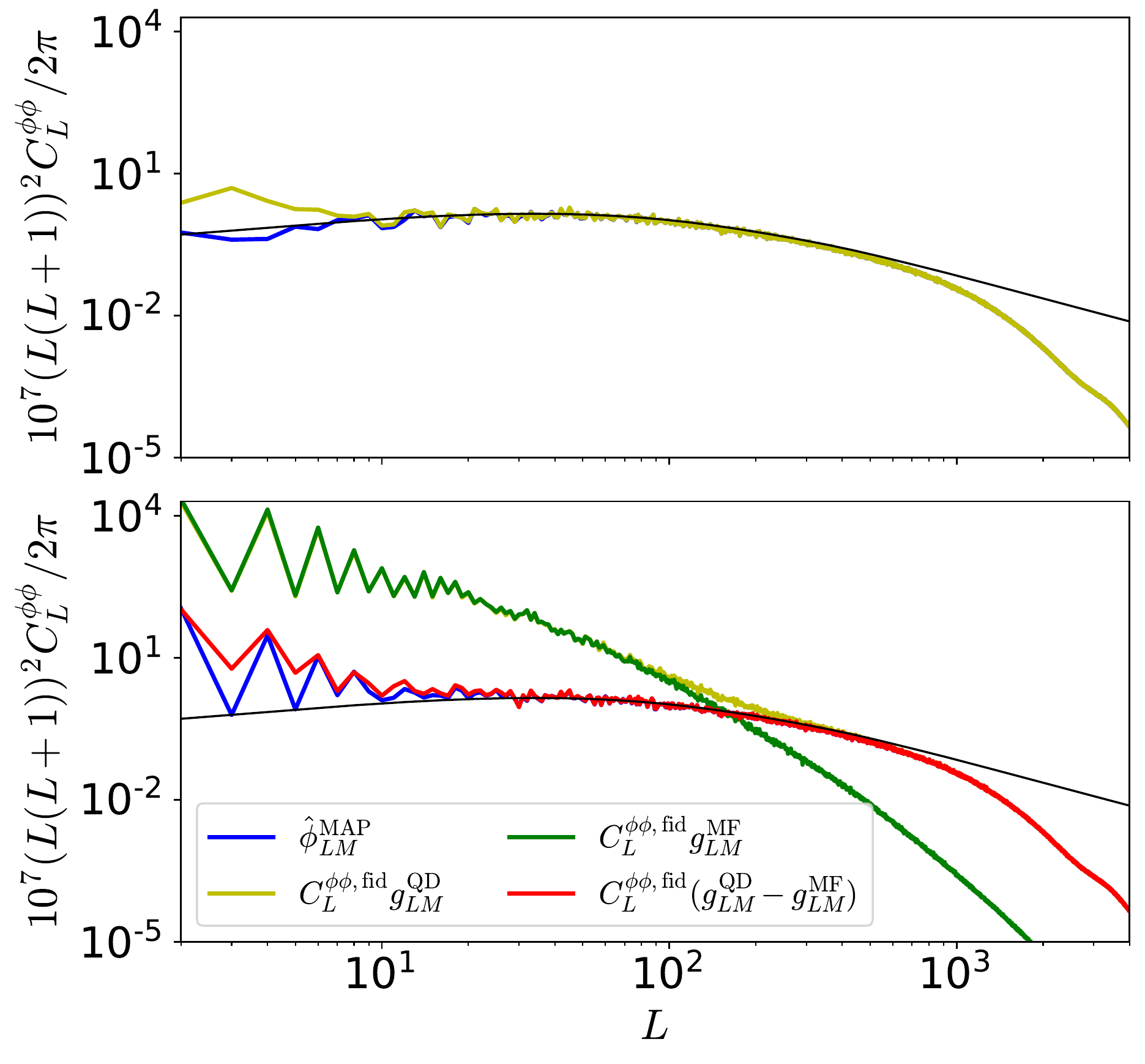}
    \caption[]{Power spectra of the gradient terms entering the likelihood, after 50 iterations for one simulation. Top panel shows the idealized full-sky reconstruction, bottom panel shows the masked reconstruction. The blue line shows the MAP lensing estimate, the olive line is the quadratic gradient term given in Eq.\ref{eq:QDgradient}, the green line is the mean-field term, here taken as the QE mean-field, estimated from a set of 320 simulations, and the red line is the difference between the quadratic and the mean-field terms, which should converge towards the MAP estimate. We see that the MAP estimate in blue contains a residual mean-field term.
    The black line shows the fiducial \cppfid. }
    \label{fig:gradients}
\end{figure}

At convergence the total gradient of Eq.~\ref{eq:gradtot} is nulled, which gives a consistency relation between the different gradient terms
\begin{equation}
    \label{eq:map_def}
    \hpMAP_{\LM} =\cppfid \left.\left( g_{\va, \LM}^{\rm QD} - g_{\va, \LM}^{\rm MF} \right)\right|_{\va = \hat \va^{\rm MAP}}
\end{equation}
where the gradients are evaluated at $\va = \hat \va^{\rm MAP}$.
In Fig.~\ref{fig:gradients} we compare the power spectra of the gradients after 50 iterations, from one CMB realization, where again the mean-field gradient is taken as the QE mean-field.

In both cases we see that the convergence equality \ref{eq:map_def} is reached, except for very large scales ($L<10$).
We will in the following consider only modes above $L>30$, as a conservative lower limit.
In practice we found that on average the lensing spectra do not change by more than 0.5\% between the iteration 20 and 50 for $L>30$, so we will use only 20 iterations for our tests with non-fiducial cases in the Section~\ref{sec:nonfidsim}. 

\begin{figure}
    \centering
    \includegraphics[width=\columnwidth]{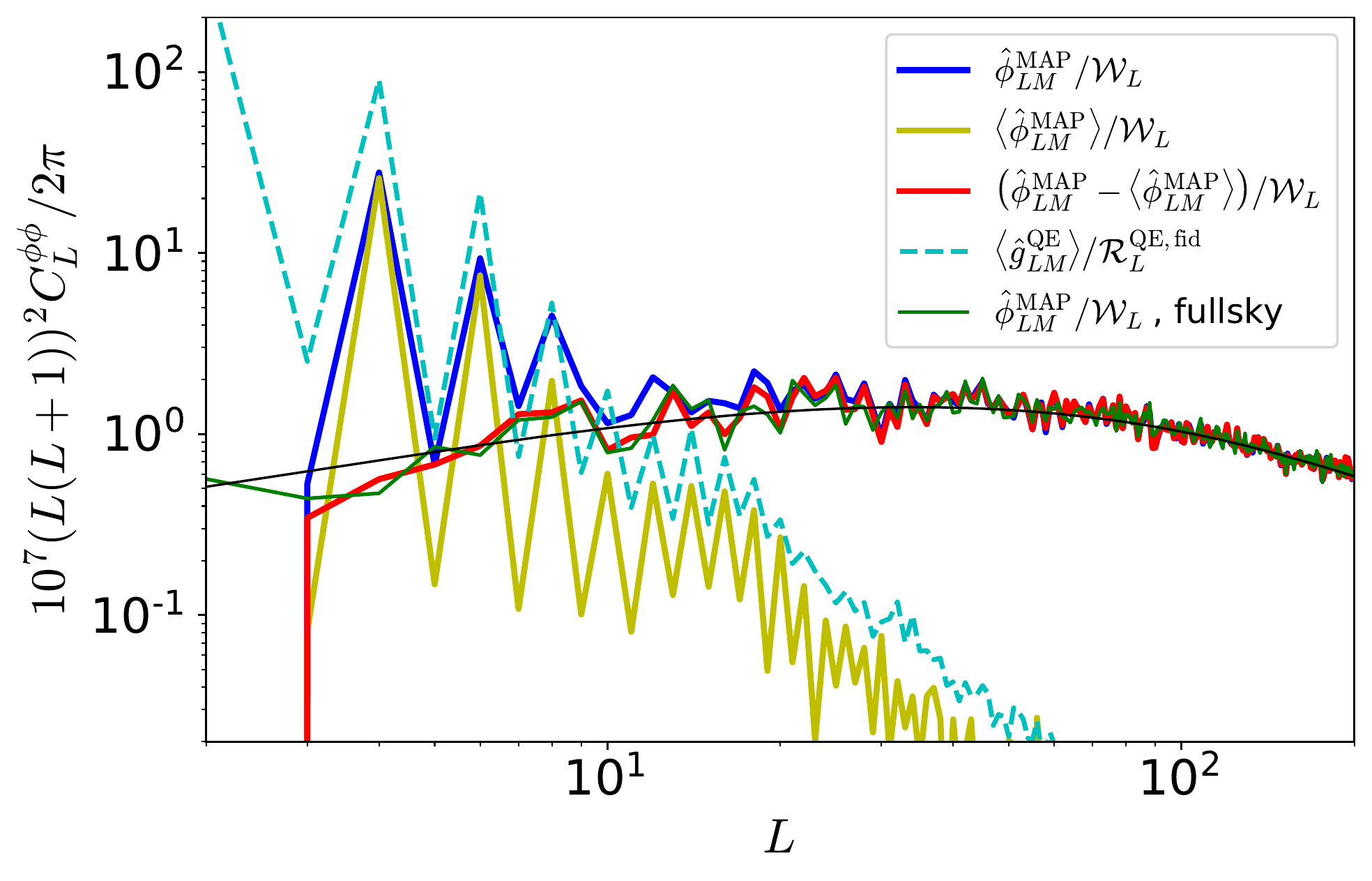}
    \caption[]{Lensing power spectrum of the MAP estimate (blue line) a  masked reconstruction for one CMB realization. The olive line shows the estimated residual mean-field, taken as the average \hpMAP from 40 simulations. The difference between the two is shown in red. For comparison we show the power spectrum for the full-sky MAP reconstruction from the same CMB realization (but without mask) in green. The QE mean-field, estimated from 320 simulations, and which is used as a template for the mean-field gradient term during iterations, is shown in dashed cyan. All these lensing fields are normalized with their respective response, corrected with the set of simulations as described in the Section~\ref{sec:norm}. The power spectra of the mean-field terms are estimated by taking the cross-correlation between two different batches of simulations to decrease the Monte-Carlo noise.}
    \label{fig:meanfieldMAP}
\end{figure}

We see in the bottom panel of Fig.~\ref{fig:gradients} that in the masked reconstruction the shape of \hpMAP at low $L$ is similar to a residual mean-field. 
In Fig.~\ref{fig:meanfieldMAP} we show an estimate of this residual mean-field using our set of simulations as introduced above. We see that subtracting this residual mean-field from the \hpMAP estimatd from one masked CMB realization, gives an estimation that is close to the \hpMAP reconstructed from the same realization but without the mask (full-sky). It then appears that it is safe to neglect the mean field sourced by the lensing field itself, and that most of the mean-field can be subtracted with these MC simulations.

\section{Normalization of the estimator}
\label{sec:norm}

Our iterative lensing field estimate $\phi^{\rm it}$ is defined by subtracting the residual mean-field from the MAP lensing field and normalizing it
\begin{equation}
    \label{eq:phi_it}
    \hat \phi^{\rm it}_{LM}= \frac{\hat \phi^{\rm MAP}_{\LM} - \av{\hat \phi^{\rm MAP}_{\LM}}}{\mathcal W^{\rm fid}_L} \; ,
\end{equation}
where $\hat  \phi^{\rm MAP}_{\LM}$ is the unormalized MAP lensing potential. The MAP mean-field $\av{\hat \phi^{\rm MAP}_{\LM}}$ is estimated as the average MAP reconstructed from a set of 40 simulations. In a first step we assume that the normalization is isotropic and is given by the Wiener-filter
\begin{equation}
    \label{eq:WFfid}
    \mathcal W^{\rm fid}_L = \frac{\cppfid}{\cppfid + 1/\mathcal R_L^{\rm MAP, fid}} \; .
\end{equation}
The response $\mathcal R_L^{\rm MAP, fid}$ is estimated iteratively following the recipe of \cite{Legrand:2021qdu}.
Similarly, the QE lensing potential $\hat \phi^{\rm QE}_{LM}$ is defined by subtracting the mean-field from the QE gradient, and normalizing with the fiducial QE response $\mathcal R_L^{\rm QE, fid}$. Obtaining QE's is cheap numerically speaking, and our QE mean-field is estimated as the average of 320 simulations. 
\begin{equation}
    \label{eq:phi_qe}
    \hat \phi^{\rm QE}_{LM}= \frac{\hat g^{\rm QE}_{\LM} - \av{\hat g^{\rm QE}_{\LM}}}{\mathcal R^{\rm QE, fid}_L} \; ,
\end{equation}
The QE response is computed using the grad-$C_\ell$, which provide the most accurate non-perturbative estimate of the response of the CMB spectra to lensing (see Appendix C. of \cite{Lewis:2011fk} and \cite{Fabbian:2019tik}). We used the very slightly sub-optimal lensed $C_\ell$ in the $g^{\rm QE}$ weights.

Anisotropies such as the mask will make the true lensing response anisotropic. This will bias the lensing field normalization.
To estimate the correction on the normalization of our estimators, we generate pairs of simulations with the same lensing field but different realizations of the primary CMB and instrumental noise, noted $i_1, i_2$. For all pairs of simulations we estimate \hpit and \hpQE as in Eqs.~\ref{eq:phi_it} and \ref{eq:phi_qe}. We compute the cross correlation power spectrum between the lensing field estimated on the pairs  $i_1, i_2$. For the masked reconstruction we compute the pseudo power spectrum, rescaled by the unmasked sky fraction \fsky. 
This cross correlation does not contain the contractions of the CMB fields which are responsible for the \Nlzero and \Nlone biases: they only contains the $C_L^{\phi \phi}$ part, and the mean field which we have subtracted already.
This allows us to get an estimate of the true estimator normalization. 
In reality the true lensing response would be a matrix $R_{LM, L'M'}$, but we make the assumption that the correction can be captured as an isotropic rescaling. In the case of the QE, this assumption was recently confirmed in detail by Ref.~\cite{Carron:2022edh}. We estimate the isotropic corrections on the normalization with
\begin{equation}
    \frac{\mathcal{R}_L^{\rm true}}{\mathcal{R}_L^{\rm fid}} = \av{\sqrt{\frac{C_L^{\hat \phi^{i_1}, \hat \phi^{i_2}}}{\cppin}}}
\end{equation}
where we average over the pairs of simulations $(i_1, i_2)$ with same lensing field $\phi^{\rm in}$, and $\mathcal R^{\rm fid}_L$ refers to both the QE response $\mathcal{R}_L^{\rm QE, \rm fid}$ and to the MAP Wiener-filter $\mathcal W_L^{\rm fid}$, depending on the estimator considered. 

\begin{figure}
    \centering
    \includegraphics[width=\columnwidth]{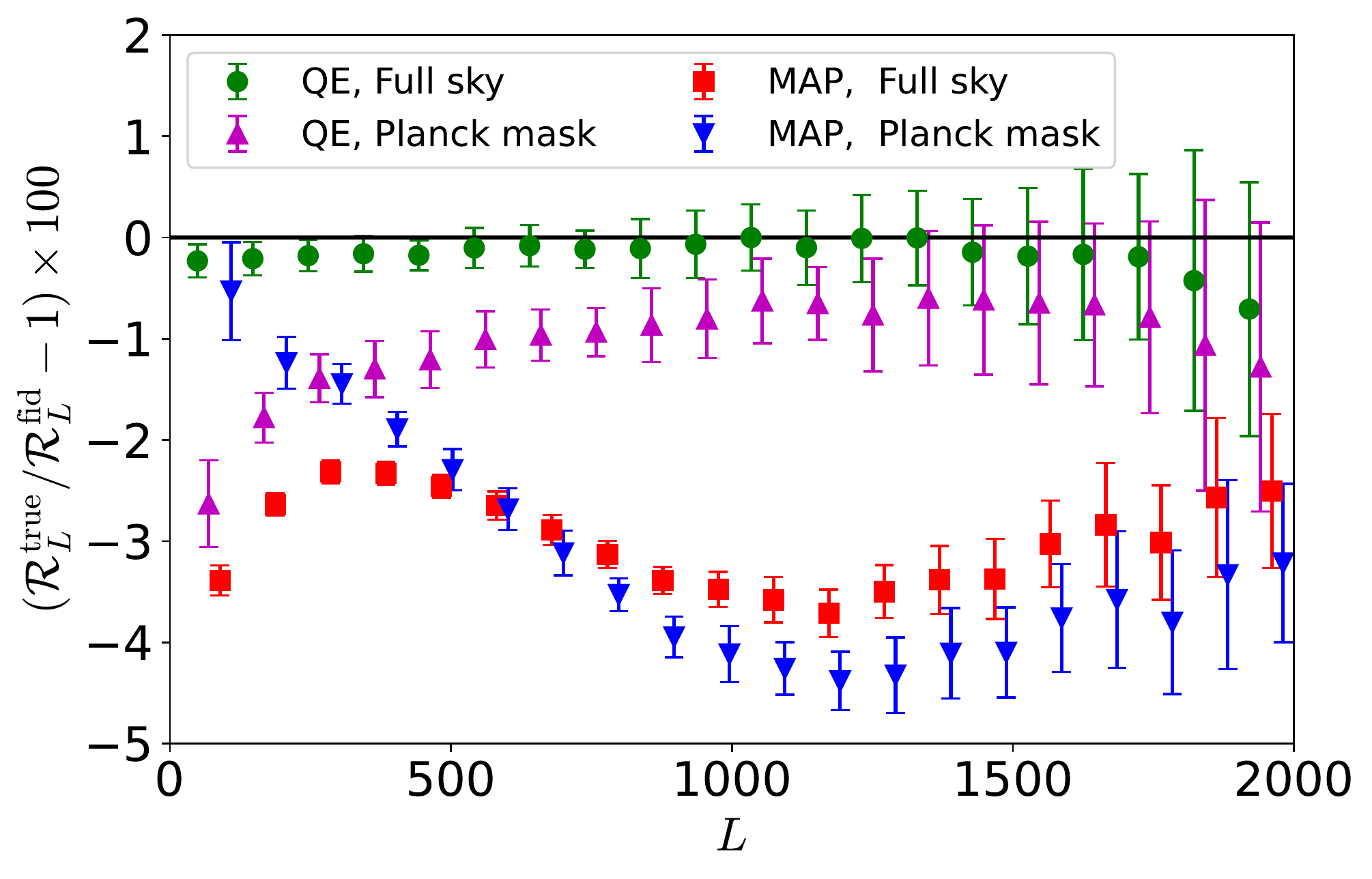}
    \caption[]{Corrections (in percent) on the lensing potential normalization, for both the QE (green circles and purple triangles) and the MAP (red squares and blue triangles). Circles and squares are for the full-sky reconstruction, while the triangles are for the masked reconstruction. This normalization bias is estimated from pairs of simulations with the same input lensing field but with different CMB realization. We show the binned average of 40 simulations, in 20 multipole bins, the error bars show the standard deviation of the simulation set in each bin. The abscisa are slightly offset for clarity.}
    \label{fig:phishuffle}
\end{figure}

We show on Fig.~\ref{fig:phishuffle} the resulting normalization bias. On the full-sky reconstruction we see that the QE has a small bias at $L<1000$, of the order of $\sim 0.2\%$.
For the MAP full-sky reconstruction, we find a correction with a maximum of $\sim 3\%$, with a similar shape as the one in our previous work \cite{Legrand:2021qdu}. However, compared to our previous work, we see a different behavior at the large scales, for $L<500$. In this previous work the correction was positive at low $L \leq 250$, while here it is negative for all multipoles. This is due to the fact that we now consider the mean-field as constant between iterations, while we previously used a perturbative mean-field approximation to estimate the $\hat \phi$-induced mean-field for each iteration. This does modify slightly the normalization of the estimator, but has no impact on the quality of the reconstruction.

On the masked sky, we recover a QE normalization correction which is similar to the one estimated for the \Planck experiment \cite{Planck:2018lbu,Carron:2022edh}, i.e. a negative $2.5\%$ bias at low multipoles, and then a correction slowing increasing (but still negative) up to around $ -1 \%$ for $L>500$.
The masked sky MAP correction has a similar shape as the MAP on the full-sky, except for the low lensing multipoles. 
% For small lenses the mask induces a loss in power similar to the QE case, but for large lenses the correction is different and switches sign.

\section{Lensing power spectrum}
\label{sec:clpp}

For a standard QE, the power spectrum of \hpQE is a four point function of the CMB fields. It contains the signal we want to measure \cpp as well as bias terms that have to be estimated and subtracted. At first order in \cpp, the spectrum results in $\av{C_L^{\hat \phi \hat \phi}} = \cpp + \Nlzero + \Nlone$, where \Nlzero is given by the disconnected part of the CMB four point function, and $\cpp + \Nlone$ is the connected part at first order in \cpp. We neglect higher order terms in this analysis, such as the $N_L^{(3/2)}$ bias due to both the non-Gaussian statistics of the density field and the post-Born lensing \cite{Fabbian:2017wfp,Fabbian:2019tik}, which is tiny for the QE in polarization.
The estimated iterative power spectrum may also be written with the above equation, with the difference that the exact correspondence of the bias terms to pieces of the CMB four-point function is lost. The \Nlzero and \Nlone biases are obtained from the iterative recipe described in \cite{Legrand:2021qdu}.

Our lensing power spectrum estimate of the QE follows very closely the \emph{Planck} reconstructions\cite{Planck:2018lbu,Carron:2022eyg}
\begin{equation}
    \label{eq:QEspectrum}
    \hat C_L^{\phi\phi, \mathrm{QE}} = C_L^{\hpQE, \hpQE} - \RDNLZERO  -\MCNLONE \;. 
\end{equation}
which we review briefly now for completeness.

We use the realization-dependent estimate of the \Nlzero bias, noted \RDNLZERO \cite{Planck:2018lbu}. This combines QE estimated from the data map and from simulations, and is robust at first order in the mismatch between the fiducial (assumed for the reconstruction) and the true CMB spectra.
Let us consider a CMB map, noted $d$, with unknown cosmology and unknown noise. We reconstruct the lensing field with a QE, using fiducial ingredients in the filtering and weights. The \RDNLZERO corresponding to this map is estimated from a set of simulations $s_i$, generated using the fiducial CMB spectra. We reconstruct the QE with a different map on each leg of Eq.~\ref{eq:QEgradient}. The first combination has one leg on the data map $d$, and another leg on a simulated fiducial map $s_i$.
The auto power spectrum of this QE is noted $\hat C_L^{d s_i}$. We then compute QE where each leg is on a different simulation $s_i$ and $s_j$, with $i \neq j$. The auto power spectrum of this QE is noted $\hat C_L^{ s_i s_j}$. In case of masking we compute the pseudo spectra, rescaled by \fsky. Because the CMB and lensing field realizations are independent, there is no mean-field contribution to subtract. The \RDNLZERO is given by combining the average of these spectra over the set of simulations
\begin{equation}
    \RDNLZERO = \frac{1}{\left(\mathcal{R}_L^{\rm QE}\right)^2} \left<4 \hat C_L^{d s_i}- 2 \hat C_L^{s_i s_j}\right>_{i\neq j} \;,
\end{equation}
with $\mathcal{R}_L^{\rm QE}$ the QE response, which has been corrected as described in the Section~\ref{sec:norm}.
This combination of data and simulations is insensitive at first order in the mismatch between the true ingredients and the fiducial assumptions.

The \MCNLONE \cite{Story:2014hni} is estimated from pairs of simulations with different CMB fields but with the same lensing potential, which we denote $s_i$ and $s'_i$. We reconstruct the QE lensing field with one leg on each simulation of the pair, and the auto power spectrum is noted $\hat C_L^{s_i s'_i}$ . The \MCNLONE is given by
\begin{equation}
    \MCNLONE =  \frac{2}{\left(\mathcal{R}_L^{\rm QE}\right)^2} \left<\hat C_L^{s_i s'_i} - \hat C_L^{s_i s_j}\right>_{i\neq j} \;.
\end{equation}
Finally, to cancel the Monte-Carlo noise due to the finite number of simulations, we split our mean-field estimate in two different batches, thus obtaining two mean-field estimates \pMFone and \pMFtwo. On the RHS of \eqref{eq:QEspectrum} we subtract a different mean-field estimate on each leg of the power spectrum. 

We now turn to the power spectrum built from the iterative, optimal lensing map. It is given by
\begin{equation}
    \label{eq:MAPspectrum}
    \hat C_L^{\phi\phi, \mathrm{it}} = C_L^{\hpit, \hpit} - \RDNLZERO  - \Nlone \;.
\end{equation}
Similarly to the QE, we split the mean-field estimate in two batches and subtract a different estimate on each leg of the power spectrum to reduce the Monte-Carlo noise.
The MAP \RDNLZERO is defined as follows: from a CMB map with unknown cosmology and noise level, noted $d$ we estimate the MAP lensing potential \hpMAP. We then generate a set of simulations noted $\mathfrak{s}_i$, with independent unlensed CMB realization and instrumental noise but all lensed by our lensing estimate \hpMAP.
We then compute the quadratic gradient term from Eq.~\ref{eq:QDgradient}, evaluated at $\va=\hat \va^{\rm MAP}$, where one leg is the simulation $\mathfrak{s}_i$ and on another leg a simulation $\mathfrak{s}_j$, with $i \neq j$.
The auto power spectrum of this gradient is noted $\hat C_L^{\mathfrak{s}_i \mathfrak{s}_j}$. We also estimate the gradient with one leg on the map $d$ and another leg on the simulation $\mathfrak{s}_i$, and its auto power spectum is noted  $\hat C_L^{d \mathfrak{s}_i}$.
The MAP \RDNLZERO is then estimated by combining the average of these power spectra over the set of simulations
\begin{equation}
    \RDNLZERO = \frac{1}{\left(\mathcal{R}_L^{\rm MAP}\right)^2} \left<4 \hat C_L^{d \mathfrak{s}_i} - 2 \hat C_L^{\mathfrak{s}_i \mathfrak{s}_j}\right>_{i\neq j} \;,
\end{equation}
where $\mathcal{R}_L^{\rm MAP}$ is the fiducial MAP response. 

Note that this $\RDNLZERO$ construction is done only once, at the final step of the iteration process. We do not attempt estimation of the lensing power spectrum from partially converged MAP lensing reconstructions.

\begin{figure}
    \centering
    \includegraphics[width=\columnwidth]{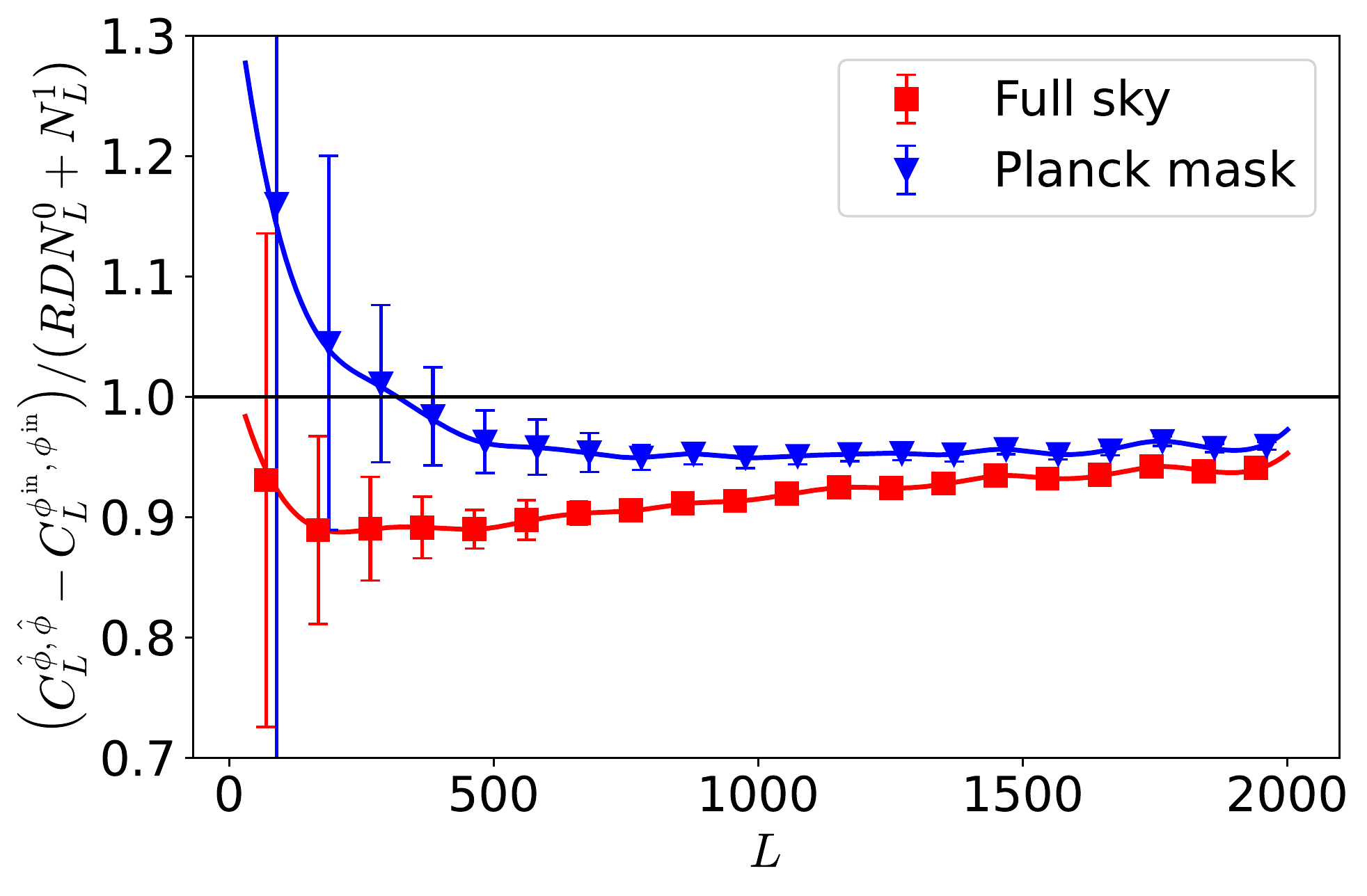}
    \caption{Correction on the response of the \RDNLZERO and \Nlone bias of the MAP. Red squares show the full-sky reconstruction and blue triangles show the masked reconstruction. This correction is estimated from a set of 40 simulations, the error bars show the variance in each bin. The bins abscissa are slightly offset for clarity. The lines show the interpolated values that we use as the response correction. The abscisa are slightly offset for clarity.}
    \label{fig:mapresp_corr}
\end{figure}

As we see here, the MAP \RDNLZERO and \Nlone are normalized with the fiducial MAP response $\mathcal{R}_L^{\rm MAP}$, while the normalization of $\hpit$ is the Wiener-filter $\mathcal W_L$ (see Eq.~\ref{eq:phi_it}). 
Both normalizations are subject to modelling errors. Hence we must estimate a second correction for the MAP response of the bias terms, which is different to the correction we estimated already for the Wiener-filter.
We estimate this correction response from a set of fiducial simulations. We estimate \hpit and the corresponding \RDNLZERO for each simulation. The iterative lensing spectrum \hpit is normalized with the corrected Wiener-filter as described in Section~\ref{sec:norm}, and the \RDNLZERO and \Nlone biases are normalized with the fiducial response. We then estimate the response correction of the bias terms with
\begin{equation}
    \label{eq:maprespcorr}
    \left(\frac{\mathcal{R}^{\rm fid}_L}{\mathcal R_L^{\rm true}}\right)^2 = \left< \frac{C_L^{\hpit \hpit} - \cppin }{\RDNLZERO + \Nlone} \right> \;.
\end{equation}
We show this correction on Fig.~\ref{fig:mapresp_corr}, for both the full-sky and the masked sky reconstructions. 
We see that this correction is almost constant, especially for the full-sky case. For the masked reconstruction, we see the impact of the mask for multipoles $L<500$, similarly as the impact we saw on the Wiener-filter in Fig.~\ref{fig:phishuffle}. We perform a cubic spline interpolation on the binned averaged points to get the effective correction that we use in the following.

\section{Non fiducial scenarios}
\label{sec:nonfidsim}

\begin{figure}
    \centering
    \includegraphics[width=\columnwidth]{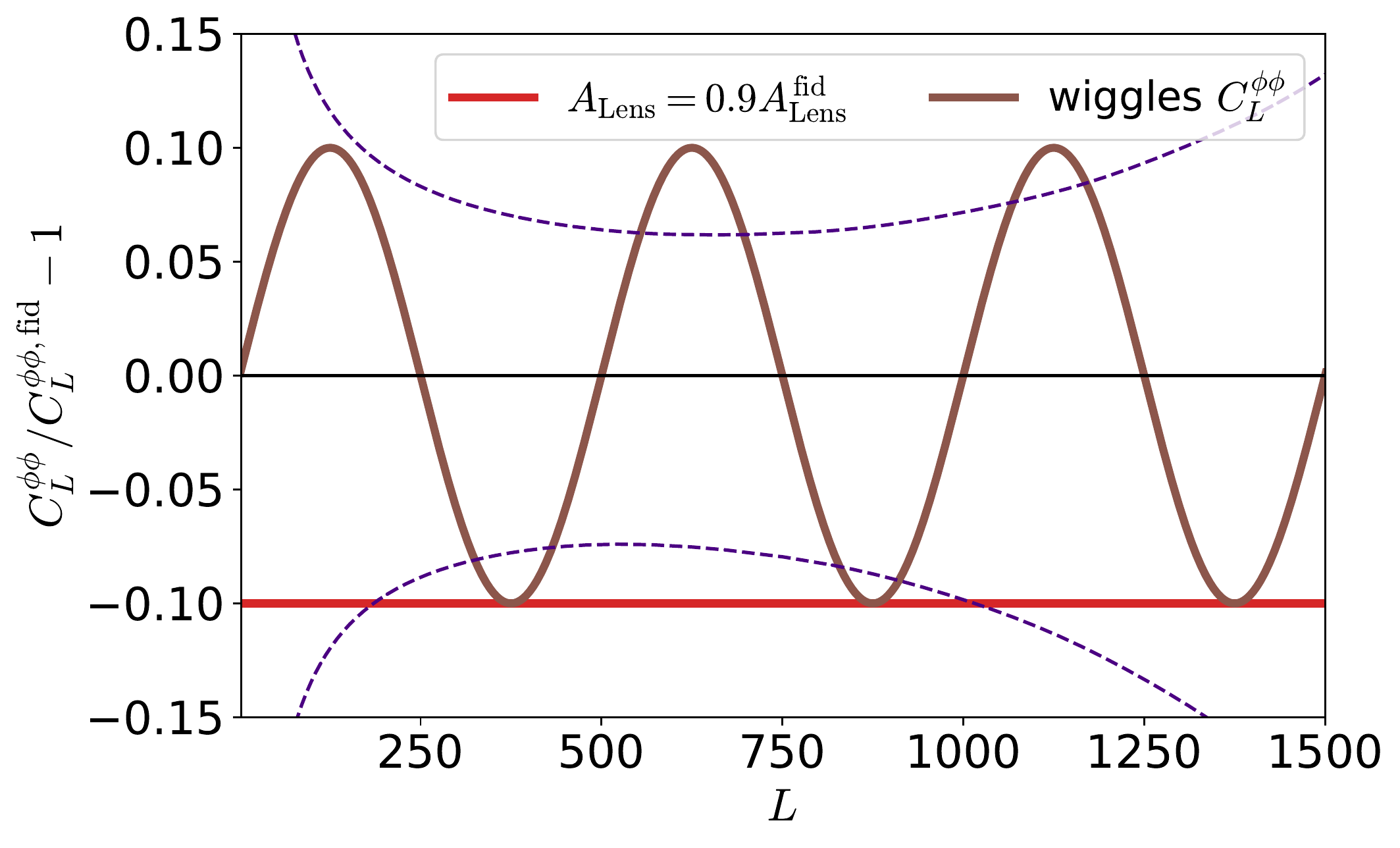}
    \caption{Non fiducial lensing power spectrum with 10\% lower amplitude (thick red line) and with the wiggles (thick brown line). The dashed purple lines show the Gaussian covariance, for the sky fraction of 67\% considered here.}
    \label{fig:nonfids_cpp}
\end{figure}

\begin{figure}
    \centering
    \includegraphics[width=\columnwidth]{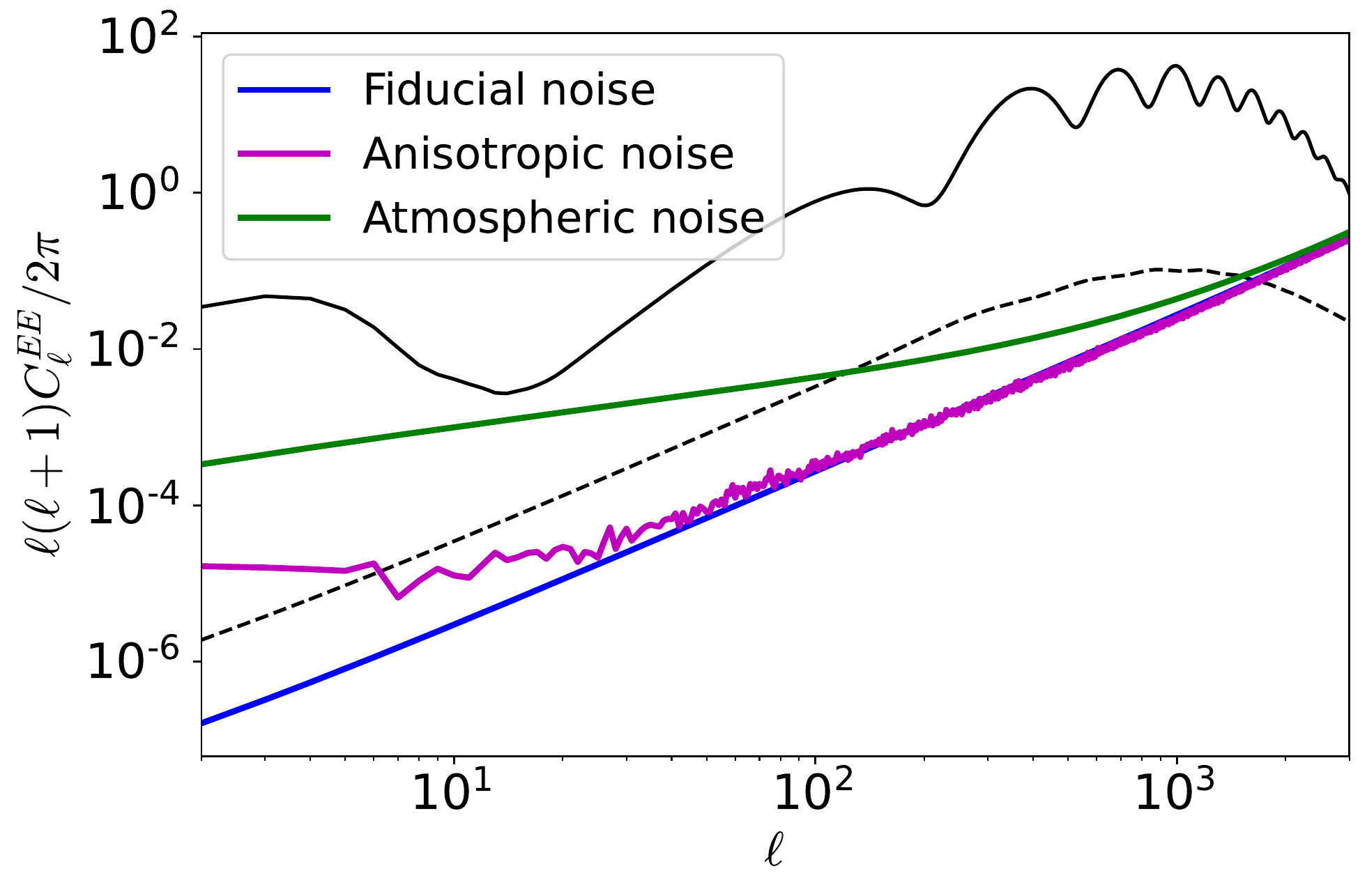}
    \caption{Polarization noise spectrum for the fiducial scenario (thick blue line), for the large atmospheric noise (thick green line) and for the anisotropic noise maps (thick magenta line). For comparison we show the lensed EE and BB spectra as the plain and dashed black lines.}
    \label{fig:nonfids_cmbs}
\end{figure}

\begin{figure*}
    \centering
    \includegraphics[width=\textwidth]{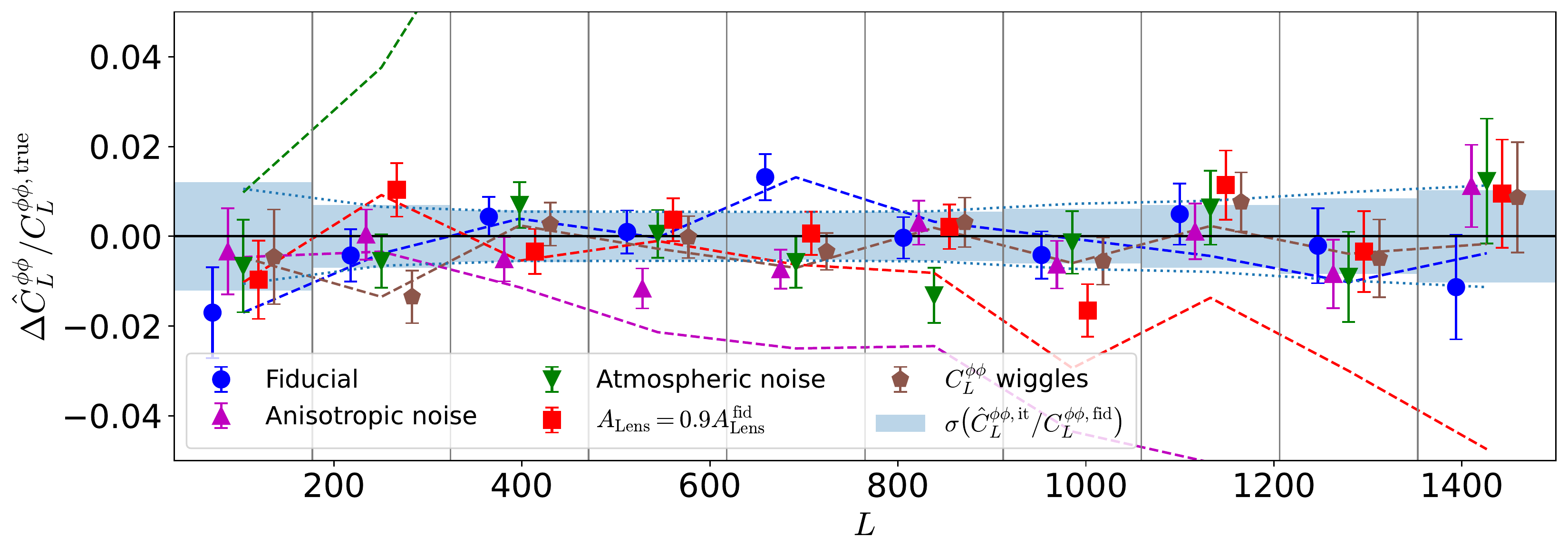}
    \caption{Residual bias on the MAP lensing power spectrum, reconstructed on the masked sky, in 10 multipole bins. Dashed lines show the residual when debiasing with the fiducial \Nlzero bias while points with error bars are obtained with the \RDNLZERO debiaser. In all cases the residuals are compared to the lensing spectrum used to generate the simulations, noted \cpptrue. We debias using the true \Nlone, computed with \cpptrue and the true lensed CMB spectra and noise levels. We show three non-fiducial cases: when the simulation contains a strong atmospheric noise (green), when the noise is highly anisotropic (magenta), when the true lensing amplitude is ten percent lower than the fiducial (red), and when the lensing power spectrum contains non-physical wiggles (brown). Error bars show the variance of the spectra in the bin, while blue boxes show the Gaussian covariance (including the cosmic variance) in each bin, estimated with the fiducial lensing spectrum. The thin dotted blue lines show the diagonal of the covariance estimated from our set of 40 fiducial simulations. Vertical dashed lines show the bins edges. }
    \label{fig:rdn0}
\end{figure*}

\begin{figure}
    \centering
    \includegraphics[width=\columnwidth]{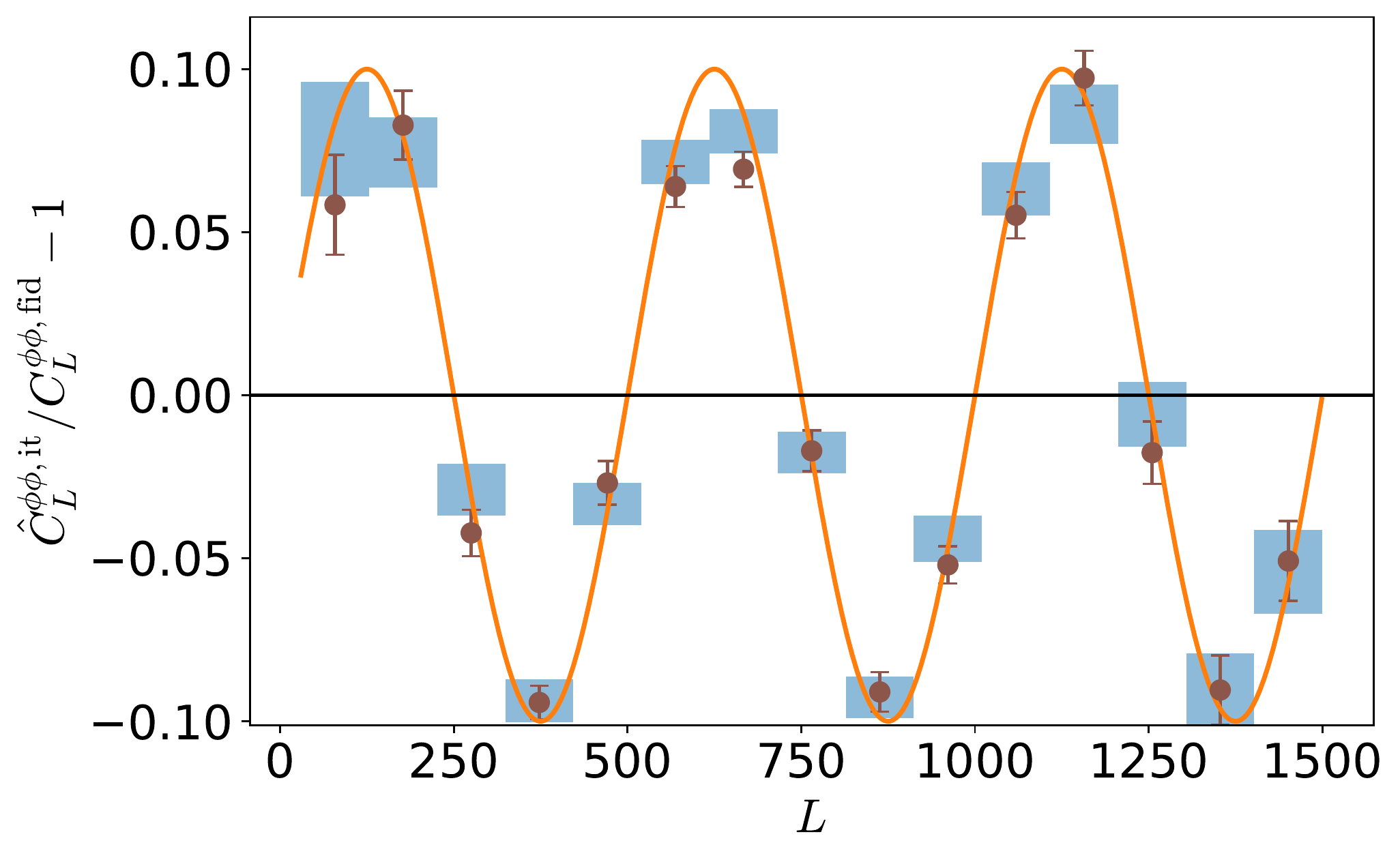}
    \caption{Ratio of the debiased iterative lensing power spectrum over the fiducial lensing power spectrum, reconstructed for the the simulation with the wiggles. Brown dots show this residual in 15 multipole bins, the error bars are the variance in each bin. The fiducial wiggles are shown in orange, and the blue boxes show the mean and Gaussian covariance of the wiggles power spectrum in each bin.}
    \label{fig:wig}
\end{figure}

\subsection{Simulations}

We now test the validity of our lensing power spectrum estimation by performing reconstructions on simulations whose signal or noise inputs are strongly different to the fiducial ingredients used in the reconstruction pipeline. 
Two cases we discuss in this section tweak the lensing signal: one where the lensing power spectrum contains non-physical wiggles, and one where the lensing power spectrum amplitude is lowered by $10\%$. Both test cases show deviations deliberately exaggerated compared to current constraints or expectations. The former tests the importance of the non-diagonal mode-coupling on the reconstruction (which are encapsulated in a diagonal multiplicative correction, as for current QE analyses), while the later takes its motivations in the main science case of the lensing spectrum being a precise measure of the amplitude of large-scale structure formation. We simulate the wiggles by multiplying the lensing power spectrum by $(1+0.1\sin(2\pi\frac{L}{500}))$. The input lensing spectra, used to generate the simulations, are shown in Fig.~\ref{fig:nonfids_cpp}. 

We also test in two ways mischaracterization of the data noise maps. In one case, we add large atmospheric noise to the data, but do not account for it in any way in the fiducial model used for the reconstruction. We model the atmospheric noise as Gaussian, multiplying the white noise $EE$ and $BB$ spectra by $\left(1+\left(\frac{\ell+1}{700}\right)^{-1.4}\right)$, which results in a noise spectrum crudely in line with expectations for the CMB-S4 large aperture telescopes. In the second case, we introduce non-isotropies and non-Gaussianities in the data maps. To do this we use a noise map from the FFP10 \Planck simulation suite. This noise map is taken from a full end-to-end simulation\footnote{\url{https://wiki.cosmos.esa.int/planck-legacy-archive/index.php/Simulation_data}}, from which the estimated CMB and foreground signals have been subtracted. We take the 143 GHz channel polarization maps, and rescale them by a constant factor so that their variance is equal to $\sqrt{2}\, \rm \mu K\: arcmin$. This rescaled map is shown in Fig.~\ref{fig:anisonoise}. The spectra of these simulations are shown in Fig.~\ref{fig:nonfids_cmbs}.

We also generate a simulation in the fiducial cosmology, independent from the simulations used to estimate the mean-field and the response corrections, for comparison.

\begin{table}
    \centering
        \begin{tabular}{ l | c | c | c | c}
            Simulation \chis & \Nlzero  &  \RDNLZERO & true \Nlone & $\hat \Cov$ \\
            \hline
            \hline
            Fiducial                                     & 0.89 & 0.96  & 0.96 & 1.02 \\
            \hline
            $C_L^{\phi\phi}$ wiggles                     & 0.79 & 0.70  & 0.70 & 0.36 \\
            $ A_{\rm lens} = 0.9 A_{\rm lens}^{\rm fid}$ & 4.66 & 1.04  & 0.86 & 1.41 \\
            Atmospheric noise                            & 2510 & 2.56  & 1.32 & 1.50 \\  
            Anisotropic noise                            & 16.6  & 0.95 & 0.93 & 1.68 \\ 
    \end{tabular}
    \caption{Values of the reduced \chis of the estimated iterative lensing power spectrum \cppit , for our set of simulations, for 20 bins in the range $L\in[30, 1500]$. The expected value of the reduced \chis is 1, with a standard deviation of 0.32. These values are obtained by debiasing the iterative power spectrum either with the fiducial \Nlzero or with the \RDNLZERO. We apply the fiducial correction of the response as described in the text. The first two columns use the fiducial \Nlone, while the third colum shows the \chis values when using the \RDNLZERO and the true \Nlone, estimated with the true spectra of the CMB maps, as shown in Fig.~\ref{fig:rdn0}. Finally the last column shows the \chis estimated using the (noisy) covariance matrix estimate from our set of 40 simulations, with the \RDNLZERO and true \Nlone.}
    \label{tab:chi2}
\end{table}

In all those non-fiducial simulations we assume the same fiducial cosmology and noise statistical model to reconstruct the lensing potential. We apply the corrections on the Wiener-filter and on the response that we estimated from our set of 40 fiducial simulations as described in the previous sections. The iterative mean-field residual is also estimated from our set of fiducial simulations.
For each simulation we estimate the corresponding \RDNLZERO bias, and correct its normalization with the bias estimated previously from the set of fiducial simulations.
We debias the iterative either with the \RDNLZERO or for comparison purposes with the fiducial analytical \Nlzero, which will be somewhat wrong by construction of the tests. For a fair comparison, the \Nlzero response is also  recalibrated such that it gives unbiased result on the fiducial model. This correction is estimated as in Eq.~\ref{eq:maprespcorr}, where we use the fiducial \Nlzero instead of \RDNLZERO in the denominator.
The \RDNLZERO is estimated using a set of 16 simulations, following the procedure defined in Section~\ref{sec:clpp}. This \RDNLZERO estimate still contains some level of MC noise, which is however sufficiently low for our purposes.

\subsection{Results}
 
We show on Fig.~\ref{fig:rdn0} the residual biases on the iterative lensing power spectrum
\begin{equation}
    \label{eq:dcpp}
    \Delta \hat C_L^{\phi\phi}= \hat C_L^{\phi\phi, \rm it} - \cpptrue \;, 
\end{equation}
where \cpptrue is the lensing power spectrum used to generate the simulation considered. We debias with the true \Nlone bias, i.e. the one estimated with the true lensing and CMB spectra of the map.

We show the residual bias up to $L_{\rm max} = 1500$. As showed in \cite{Legrand:2021qdu}, the signal to noise ratio of the CMB lensing power spectrum saturates above this value for a CMB-S4 survey. We also found that the MC noise of the normalization and response correction for the MAP was high above this threshold, owing to the small signal. The analysis could be extended to higher multipoles by performing a larger set of MC simulations, but the improvement in terms of signal is expected to be small.

We show in Table.~\ref{tab:chi2} the reduced \chis values of our debiased power spectra, Eq.~\eqref{eq:dcpp}, which gives an estimate of the goodness of fit of the true input cosmology to $\hat C_L^{\phi\phi, \rm it}$.
This reduced \chis is defined as
\begin{equation}
    \chi^2 = \frac{1}{N_{\rm bin}} \sum_{L_i L_j}\Delta \hat C^{\phi\phi}_{L_i} \Cov^{-1}_{L_i L_j} \Delta \hat C^{\phi\phi}_{L_j} \; ,
\end{equation}
with $N_{\rm bin}$ the number of bins considered and $L_i$, $L_j$ the binned multipole indices.
We estimate the \chis with a Gaussian analytical covariance, computed in the fiducial settings, given by
\begin{equation}
    \Cov_{L L'} = \delta_{LL'} \frac{2}{(2L+1) \fsky} \left(\cppfid + \Nlzero\right)^2 \;.
\end{equation}
We also estimate the binned covariance matrix from our set of 40 fiducial simulations. The empirical covariance matrix is noisy, and we correct for the bias in the inverse covariance estimate following \cite{Hartlap:2006kj}.
We consider 20 bins between 30 and 1500 to compute the reduced \chis, this gives an expected standard deviation of $\sigma_{\chi^2} = 0.32$, for an expected value of 1.
We compare the goodness of fit when debiasing with the fiducial \Nlzero and \Nlone, when using the \RDNLZERO, and when using the true \Nlone bias as well.

We detail below the results for each scenario.
\begin{description}
\item[Fiducial] 
The fiducial simulation is used for comparison: it is independent from our set of 40 simulations used to estimate the response corrections. We see that the \Nlzero debiasing is already a good fit (as expected in this case), and the \RDNLZERO debiasing is similarly consistent.
    \item[Wiggles] 
Using the fiducial \Nlzero bias in the scenario with the wiggles in the lensing power spectrum also gives a good fit, and the \RDNLZERO does not improve.
We found that the \Nlzero bias computed with the true lensing power spectrum is in fact almost identical (a relative difference less than 0.1\%) to the fiducial \Nlzero, and similarly for the \Nlone bias. 
The wiggles do not give a strong bias in the \Nlzero and \Nlone bias of the iterative estimator. We conclude that this scenario is mainly a test of the accuracy of the signal reconstruction rather than a test of the \RDNLZERO bias. In particular, this shows that the lensing power spectrum used as a prior does not bias the reconstructed power spectrum in this test case (which is fairly specific, as has roughly same integrated total power). We show the ratio of the reconstructed lensing spectrum over the fiducial one in the Fig.~\ref{fig:wig}. We see that the reconstructed lensing power spectrum consistently recovers the shape of the wiggles contained in the true lensing spectrum.
\item[Lensing amplitude]
The scenario with the 10\% lower lensing amplitude gives however a biased spectrum when we consider the fiducial \Nlzero, with a $11\sigma$ tension in the \chis value. Indeed, the lower amplitude of lensing power results in a lower $B$-mode power and lower reconstruction noise on the predicted delensed spectra, and the fiducial \Nlzero is biased high by around $\sim3\%$ compared to the \Nlzero computed with the true lensing spectrum. Using the \RDNLZERO recovers a much more consistent residual lensing power spectrum. Debiasing with the true \Nlone, as one would do when performing a cosmological analysis close to the best-fit, further decreases the \chis value.
\item[Atmospheric noise]
For the simulation with atmospheric noise we see that the fiducial \Nlzero would be totally off, and the \RDNLZERO debiaser greatly reduces it. But even when using the \RDNLZERO debiaser, the \chis has a high deviation of $5\sigma$ from the expected value. This is due to a bias in the \Nlone estimate. As we already noted in our previous work \cite{Legrand:2021qdu}, the iterative \Nlone, unlike the QE \Nlone, has a dependency on the data noise, since the amount of noise impacts the achievable amount of delensing. Using the true noise spectra in the \Nlone estimate improves the fit, and brings the \chis to a consistent value.
\item[Anisotropic noise]
The simulation with the anisotropic noise might be the most relevant test of the \RDNLZERO debiaser. Indeed, in this highly anisotropic case, the fiducial \Nlzero is totally off, even if the variance of the maps is equal to the fiducial variance. The \chis has a $48\sigma$ tension from the expected value. However, using the \RDNLZERO completely resolves the tension. 

\end{description}

The \chis values computed with the covariance matrix estimated from our simulations is roughly in agreement with the fiducial covariance matrix. As we only used 40 simulations, for 20 multipole bins, this covariance matrix is noisy. Following \cite{Taylor:2012kz}, this gives a $17\%$ error on the uncertainty, in that case the variance of the \chis is of $0.32 (1 \pm 0.17)$, which reduces a bit the tension with the \chis computed with the estimated covariance.
We note however in Fig.~\ref{fig:rdn0} that the diagonal of the covariance estimated from the set of simulations (dashed blue lines) matches the Gaussian covariance (blue boxes). Because we used only 40 simulations, we cannot conclude on the non diagonal terms of the covariance, especially since we expect the mask to create non diagonal correlations. However, as shown in \cite{Legrand:2021qdu} in the full sky case, the MAP lensing power spectrum covariance matrix is diagonal. This shows that the forecasts usually performed with the Gaussian covariance should be close to what we will obtain from the lensing reconstruction.

These different scenarios confirm that our corrections, estimated from the fiducial simulations, essentially depends on the fiducial ingredients only, and we are able to recover unbiased lensing power spectrum estimates even when the true inputs vary from these in fairly extreme ways. We have also seen that the realization-dependent noise debiaser \RDNLZERO is critical in doing so for realistic noise maps.

\section{Conclusions}

The efficient extraction of the CMB lensing power spectrum signal from low noise polarization CMB data requires more elaborated tools than quadratic estimators. In this work, we have pursued the approach initiated in~\cite{Legrand:2021qdu}, that uses the spectrum of the optimal lensing mass map, built assuming a fiducial cosmological model, and characterizing properly its response to the signal and noise biases, working in close analogy to quadratic estimator theory. We have investigated the response and biases of the estimator to masking and to more realistic, non-isotropic noise maps.

Our main conclusion is that following the QE approach is still very satisfactory in this realistic setting. The mean-field impacting the optimal lensing map can be estimated in economical manner in basically the same way. At the level of the lensing spectrum band-powers, the mask induces a correction of similar size than for the QE. In both cases, it is very challenging to predict the form of this correction analytically, but it can be calibrated using simulations without much difficulties. Since the lensing power spectrum is very smooth in all realistic cosmological models, the non-diagonal mixing of the modes is of minor relevance, making a diagonal correction good enough.

Also, for the scenarios we considered in a CMB-S4 like configuration, the realization-dependent debiaser \RDNLZERO we introduced in \cite{Legrand:2021qdu} remains a very robust way to remove the leading noise term to the spectrum estimate. We see this as an important feature of the estimator, since the precise statistics of the instrumental noise are sometimes only very crudely understood. We showed that the \RDNLZERO debiaser is effective on reconstructions from maps where the noise is given by the complex end-to-end noise simulations of the \emph{Planck} satellite.

We also showed that the covariance matrix estimated from our set of simulations is close to the naive analytical Gaussian covariance, with the number of modes reduced by the sky fraction, and $\Nlzero$ obtained with an analytical iterative procedure building upon a first proposal from~\cite{Smith:2010gu}. This validates the standard simple forecasts performed with the analytical covariance. Conversely, this supports the idea that the estimator is optimal, in the sense of capturing all of the available signal to noise. The CMB-S4 \cite{CMB-S4:2016ple} survey is expected to provide a $4\sigma$ detection of the neutrino mass (in minimal mass scenario), when combining primary CMB with BAO and CMB lensing measurements. The results presented here show that these forecasts should be reached with our optimal lensing estimator on masked maps.

From a practical point of view, the costliest step, numerically speaking, is the production of the optimal lensing map. All the other steps have a similar counterpart in the QE approach. This construction is performed only once per map or simulation to analyse, providing a massive simplification compared for example to sampling-based or iterative methods. In fact, we view this estimator as likely the most economical, yet optimal, estimator of the lensing spectrum. Together with the realization-dependent debiasing, the estimator also appears reasonably robust, and we plan in future work to use it on actual CMB lensing data.

Probably the main difference to a state-of-the-art QE analysis on a masked sky is that we had to introduce and calibrate two independent corrections, instead of one. The Wiener-filtered nature of the optimal lensing map (which is obtained from maximizing a posterior instead of a likelihood function) causes in our implementation the signal and noise part of the spectrum to react differently to masking. The QE case is simpler, both terms sharing the same normalization. Nevertheless, these corrections depend essentially only on the fiducial cosmology chosen for the lensing map reconstruction, which allows simple calibration of both of them, and we could recover the expected results in all of our test cases. It is possible that a more detailed understanding of the relation between the Wiener-filter and the iterative response could provide clues to the origin of these corrections and possibly reduce them to a single one. An analytical understanding seems challenging at the present time however.

We neglected the impact of polarized foregrounds in our reconstruction. We expect those to be less important than in temperature, but a thorough analysis which should also include foreground cleaning is beyond the scope of this paper and left for future work. We also neglected the non-Gaussianity of the lensing field as well as post-Born lensing, which leave a very small signature on the polarized quadratic estimators. This is currently under study and will be reported elsewhere.

\begin{acknowledgments}
We thank Antony Lewis for insightful discussions and comments on the draft. We acknowledge support from a SNSF Eccellenza Professorial Fellowship (No. 186879). Parts of this research used
resources of the National Energy Research Scientific Computing Center (NERSC), a U.S. Department of Energy Office of Science User Facility located at Lawrence Berkeley National Laboratory.
\end{acknowledgments}

\bibliography{cleanbib}

\end{document}